\documentclass[review,12pt]{elsarticle}

\journal{Materials Research Bulletin}

\usepackage[british]{babel}
\usepackage{hyphenat}

\usepackage{lineno,hyperref}
\modulolinenumbers[5]

\usepackage{graphicx,amssymb}
\usepackage{color}

\begin{document}

\begin{frontmatter}

\title{Revealing the local structure of CuMo$_{1-x}$W$_x$O$_4$ solid solutions by multi-edge X-ray absorption spectroscopy}


\author[ISSP]{Inga Pudza}

\author[ISSP]{Andris Anspoks}

\author[ELETTRA]{Giuliana Aquilanti}

\author[ISSP,KUMUNI]{Alexei\ Kuzmin\corref{cor1}}
\ead{a.kuzmin@cfi.lu.lv}
\cortext[cor1]{Corresponding author}

\address[ISSP]{Institute of Solid State Physics, University of Latvia, Kengaraga Street 8, LV-1063 Riga, Latvia}

\address[ELETTRA]{Elettra -- Sincrotrone Trieste S.C.p.A., Ss 14, Km 163.5, I-34149 Basovizza, Trieste, Italy}

\address[KUMUNI]{International Research Organization for Advanced Science and Technology (IROAST), Kumamoto University, 2-39-1 Kurokami, Chuo-ku, Kumamoto 860-8555, Japan}


\begin{abstract}
The effect of tungsten substitution with molybdenum  on the structure  of CuMo$_{1-x}$W$_x$O$_4$ ($x$ = 0.20, 0.30, 0.50, 0.75) solid solutions was studied by multi-edge X-ray absorption spectroscopy. The simultaneous analysis of EXAFS spectra measured at several (Cu K-edge, Mo K-edge and W L$_3$-edge) absorption edges was performed by the reverse Monte Carlo  method taking into account  
multiple-scattering effects. The degree of distortion of the coordination shells  and its dependence on the composition were estimated from partial radial distribution functions (RDFs) $g(r)$ and bond angle distribution functions (BADFs) $f(\varphi)$. The analysis of partial RDFs suggests that the structure of solid solutions is mainly determined by the tungsten-related sublattice, while molybdenum atoms adapt to a locally distorted environment. As a result, the coordination of both tungsten and molybdenum atoms remains octahedral as in CuWO$_4$ for all the studied compositions. 
\end{abstract}

\end{frontmatter}


\newpage


\section{Introduction}

 CuMo$_{1-x}$W$_x$O$_4$ solid solutions possess many interesting properties that directly depend on their composition and are conditioned by their crystallographic structure. Under ambient conditions, the Mo$^{6+}$ and W$^{6+}$ ions have distinct (tetrahedral and octahedral, respectively) coordination by oxygen atoms in pure CuMoO$_4$ and CuWO$_4$.  Therefore, their mixing affects largely the structure of solid solutions and allows one to tailor their properties.  
Indeed, the solid solutions exist in one of the three crystal structures with triclinic $P\bar{1}$ symmetry, which are isostructural to the high-pressure phases of CuMoO$_4$ ($\alpha$, $\gamma$, III) \cite{Wiesmann1997,Yanase2013,Benchikhi2017}.

In pure CuMoO$_4$ and at low tungsten content ($x < 0.15$), a first-order phase transition between $\alpha$ and $\gamma$ phases takes place and is accompanied by a change in the coordination of  molybdenum/tungsten atoms by oxygen atoms from tetrahedral to octahedral \cite{Gaudon2007a,Robertson2015}. 
The phase transition can be induced by temperature  or pressure and is accompanied by a color change, demonstrating thermochromic or piezochromic properties \cite{Yanase2013,Gaudon2007a,Robertson2015,Rodriguez2000,Gaudon2007b,Thiry2008,Blanco2015,Morelle2019,Joseph2020,Panchal2022}. The high-temperature (low-pressure) phase is greenish, whereas the low-temperature (high-pressure) phase is brownish  \cite{Gaudon2007a,Robertson2015}. 

The pronounced chromic properties of materials based on CuMoO$_4$ are closely related to changes in their local environment under the influence of external stimuli and open up various ways of their applications extending from user-friendly temperature and pressure indicators to cost-effective ``smart'' inorganic pigments, light filters and sensors  \cite{Yanase2013,Gaudon2007a,Gaudon2007b,Thiry2008,Joseph2020,Gaudon2010}.

Nanostructured CuMoO$_4$ with a crystallite size of about 134~nm was proposed recently  for use as bifunctional catalysts for electrochemical water splitting and CO$_2$ reduction \cite{Rahmani2020}. 
Nevertheless, its electrocatalytic activity turned out to be worse than that of another molybdate Cu$_3$Mo$_2$O$_9$ due to a larger density of catalytically active sites (copper ions) in the latter caused by the unique topology  of its lattice \cite{Rahmani2020}.  At the same time, CuMoO$_4$ appears to be more promising for CO$_2$ reduction
due to the stronger adsorption of CO$_2$ species at its surface \cite{Rahmani2020}.

CuMoO$_4$-CoMoO$_4$ hybrid microspheres with hollow structure and decorated with nanorods on their surface demonstrated ultrahigh catalytic performance in hydrogen production via hydrolysis process from a hydrogen storage material as ammonia borane \cite{Feng2021}. 
The use of mixed molybdates as a catalyst reduces the energy barrier for the reaction stage determining its rate to the lowest value compared to pure molybdates \cite{Feng2021}.

A hybrid Cu$_2$O/CuMoO$_4$ nanosheet electrode with porous texture and hence a large surface area was suggested  in \cite{Du2016} for use as an asymmetric supercapacitor in tandem with an activated carbon electrode.
To reduce internal resistance, the hybrid can be directly grown on the conductive substrate as Ni foam with a 3D porous structure. The presence of MoO$_4^{2-}$ anions during the growth process facilitates the formation of nanosheets  \cite{Du2016}.

The supercapacitive behaviour was also observed in mixed CuMoO$_4$/ZnMoO$_4$ composite with nanoflower morphology and numerous pore spaces \cite{Appiagyei2021}. It is characterized by a high specific area and electrical conductivity. Compared to pure molybdates, the composite showed favorable specific capacity and stability \cite{Appiagyei2021}.

The solubility limit of tungsten in CuMo$_{1-x}$W$_x$O$_4$ solid solutions to retain both $\alpha$ and $\gamma$ structure types under ambient conditions is between $\sim$10\% and 15\% \cite{Gaudon2007a,Gaudon2008}. An increase of the tungsten content above $\sim$15\%
leads to the wolframite-type (CuMoO$_4$-III) phase with the octahedral coordination of metal ions and brownish colour  \cite{Wiesmann1997,Thiry2008,Hill2013}.
This phase does not exhibit pronounced thermochromic behaviour and has properties that are promising for possible use as photoanodes in water-splitting photoelectrochemical cells \cite{Wiesmann1997,Gaudon2007a,Hill2013,Liang2018,YANG2019195}.

The  chromic properties of CuMo$_{1-x}$W$_x$O$_4$ solid solutions are determined by 
the interband transition due to the oxygen-to-metal charge transfer  across  
the  band  gap  and the intraband d-d transition of Cu$^{2+}$ ions \cite{Gaudon2007a,Rodriguez2000,Steiner2001,Wu2020}. 
The intraband transitions contribute to absorption in the red spectral range below 1.8~eV  \cite{Gaudon2007a,Rodriguez2000,Hernandez1999}, whereas the interband  
one depends on the composition. The bandgap is $\sim$2.3~eV in pure CuWO$_4$ with distorted octahedral coordination of tungsten atoms \cite{Ruiz-Fuertes2010,Kuzmin2013cuwo4} but decreases to $\sim$2.0~eV in CuMo$_{0.65}$W$_{0.35}$O$_4$  \cite{Hill2013,Liang2018}. 
The bandgap reaches its minimum before the phase transition to the $\alpha$-phase. 
When the transition occurs, the bandgap starts to increase from  $\sim$2.2~eV in CuMo$_{0.88}$W$_{0.12}$O$_4$ \cite{Wu2020} to above $\sim$2.5~eV in $\alpha$-CuMoO$_4$ with tetrahedral coordination of molybdenum atoms \cite{Gaudon2007a,Rodriguez2000,Steiner2001,Hernandez1999}. 
Thus, the electronic structure and related properties of  CuMo$_{1-x}$W$_x$O$_4$ solid solutions depend on the local environment of metal atoms, which can be successfully probed by  X-ray absorption spectroscopy (XAS). Our previous XAS studies were dedicated to CuMoO$_4$ \cite{Jonane2018b,Jonane2018c,Jonane2019b} and CuMo$_{0.90}$W$_{0.10}$O$_4$ \cite{Jonane2020a}. Recently, we demonstrated that the resonant X-ray emission spectroscopy (RXES) at the W L$_3$-edge can be used to determine the crystal-field splitting parameter for the 5d(W)-states that allows one to distinguish between tetrahedral and octahedral coordination of tungsten atoms in CuMo$_{1-x}$W$_x$O$_4$ solid solutions \cite{Pudza2021a}.

In this study, we focus on the effect of tungsten substitution with molybdenum on the structure of CuMo$_{1-x}$W$_x$O$_4$ solid solutions for high tungsten content with $x \geq 0.20$. The low symmetry of such compounds makes the analysis of the extended X-ray absorption fine structure (EXAFS) a challenging task due to a large number of scattering paths \cite{Kuzmin2020treatment}.  To address this issue, the reverse Monte Carlo method \cite{Timoshenko2012rmc,Timoshenko2014} was used to determine the structural model of a  particular solid solution, which is  in agreement simultaneously with several experimental EXAFS spectra measured at the absorption edges of metals present in  the material.

\section{Experimental}

Polycrystalline CuMo$_{1-x}$W$_x$O$_4$ ($x$ = 0.20, 0.30, 0.50, 0.75) solid solutions  
were synthesized using solid-state reaction method and characterized in our previous studies \cite{Jonane2020a,Pudza2021a}. Briefly, stoichiometric mixtures of CuO, MoO$_3$ and WO$_3$ were heated at 650$^\circ$C in air for 8 hours and then cooled to room temperature. One set of samples with $x\leq 0.15$ was greenish, and the second set with $x \geq 0.20$ was brownish (see the inset in Fig.\ \ref{fig2}(a)). Pure CuMoO$_4$ and CuWO$_4$ were used for comparison.  

\begin{figure*}[t]
	\centering
	\includegraphics[width=0.5\linewidth]{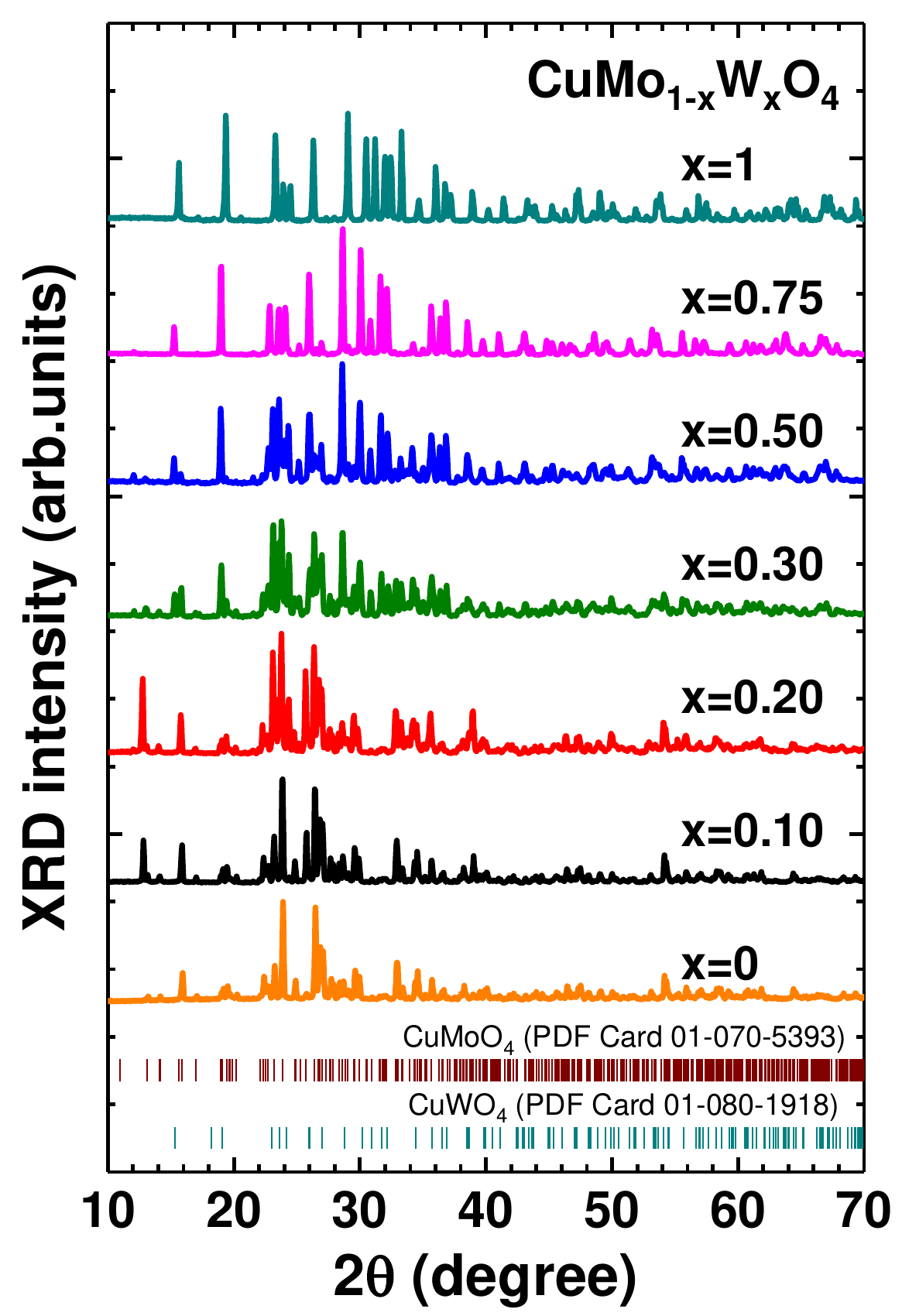}
	\caption{X-ray diffraction patterns of polycrystalline CuMo$_{1-x}$W$_x$O$_4$  solid solutions.
		The standard PDF cards of CuMoO$_4$ (PDF Card 01-070-5393) and CuWO$_4$ (PDF Card 01-080-1918) phases are shown for comparison. }
	\label{fig1}
\end{figure*}

\begin{figure*}[t]
	\centering
	\includegraphics[width=\linewidth]{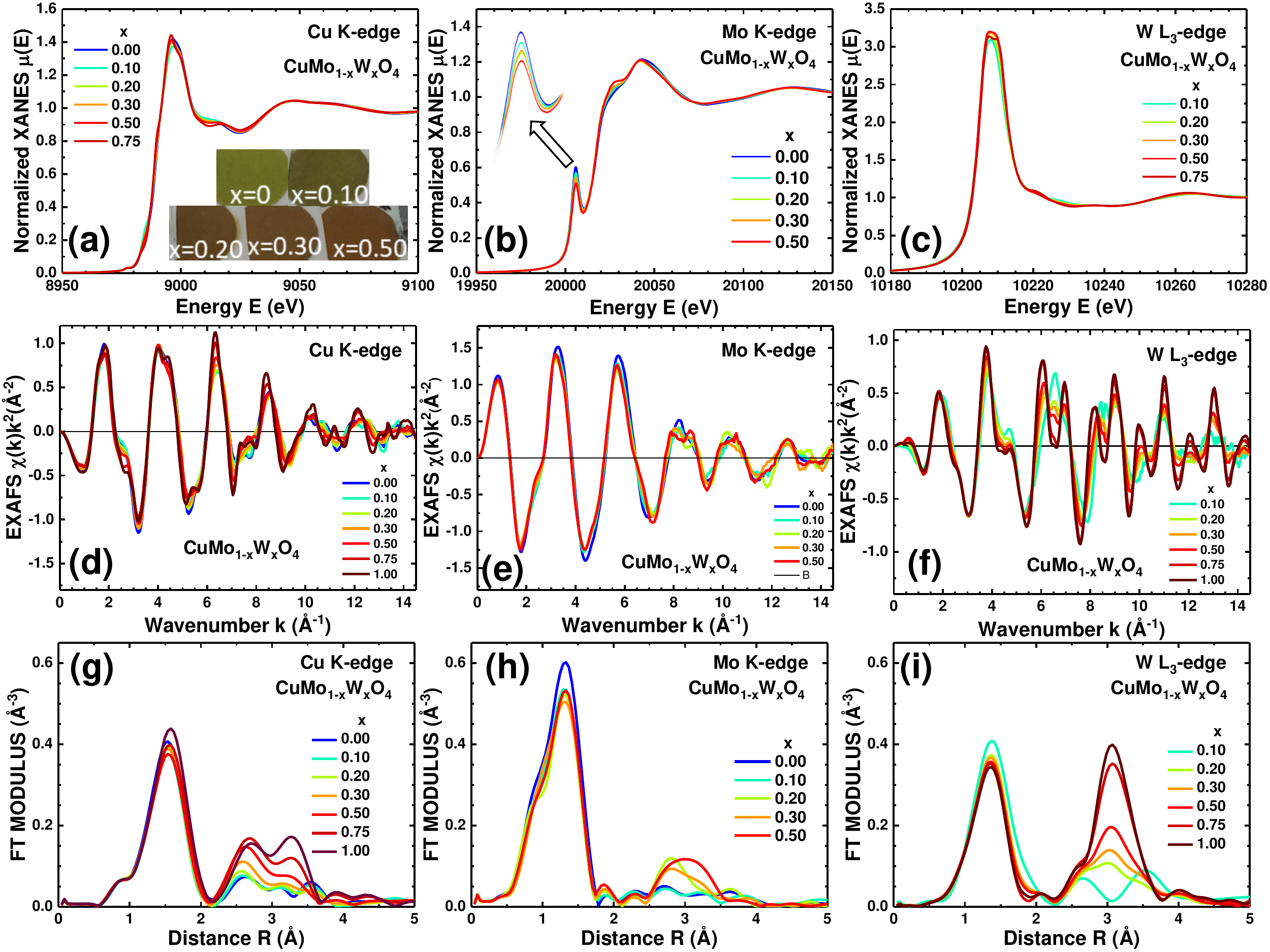}
	\caption{The experimental Cu, Mo K-edge and W L$_3$-edge normalized XANES (a-c) and EXAFS  $\chi(k)k^2$ (d-f) spectra and their Fourier transforms (g-i) for CuMo$_{1-x}$W$_x$O$_4$  solid solutions and pure CuMoO$_4$ ($x$ = 0) and CuWO$_4$ ($x$ = 1). The FTs were calculated in the $k$-space range of 2.5--14.5~\AA$^{-1}$. The selected samples are shown in the inset in (a). }
	\label{fig2}
\end{figure*}

All samples were characterized by X-ray powder diffraction. X-ray diffraction patterns of CuMo$_{1-x}$W$_x$O$_4$ powders (Fig.\ \ref{fig1}) were measured at 300~K using a Brucker D2 PHASER $\theta$-$\theta$ powder diffractometer equipped with a copper anode (Cu K$\alpha$) tube. The transition from $\alpha$-CuMoO$_4$ phase to wolframite-type CuWO$_4$ phase occurs above $x$$\geq$0.2 \cite{Pudza2021a}. 

Multi-edge X-ray absorption spectroscopy experiments of CuMo$_{1-x}$W$_x$O$_4$  samples were performed at room temperature in transmission mode at the Elettra XAFS bending-magnet beamline \cite{Elettra2009}. The spectra were collected at the Cu (8979~eV) and Mo (20000~eV) K-edges and W (10207~eV) L$_3$-edge. 
Elettra storage ring operated in a top-up mode in which frequent beam injections maintained a quasi-constant beam current for user operations. In particular, at 2.4~GeV the stored beam current is set to 160~mA and top-up occurs every 20 minutes, injecting 1~mA in 4~s to maintain the current level constant to less than 7\%.
The synchrotron radiation was monochromatized using Si(111) (for the Cu K and W L$_3$ edges) and Si(311) (for the Mo K-edge) double-crystal monochromators. 
Three ionization chambers filled with a mixture of Ar, He and N$_2$ gases were used to measure the radiation intensity before and after the sample and after the reference.
Powder samples were gently milled in an agate mortar and deposited on the Millipore membranes. Reference Cu, W and Mo metal foils were used for energy calibration. 
Each sample was measured twice with a large signal-to-noise ratio so that the experimental EXAFS spectra obtained coincide almost perfectly after multiplication by $k^2$.
Normalized X-ray absorption near edge structure (XANES) spectra, the EXAFS spectra $\chi(k)k^2$ extracted using the Athena code \cite{Ravel2005athena} and their FTs calculated with the Gaussian window using the EvAX code \cite{Timoshenko2012rmc,Timoshenko2014,evax2017} in the $k$-space range of 2.5--14.5~\AA$^{-1}$\   are shown in Fig.\ \ref{fig2}. Note that the positions of peaks in the FTs were not corrected for the phase shift, therefore, they are located at slightly shorter than crystallographic distances.

\section{Methods}

EXAFS spectra of CuMo$_{1-x}$W$_{x}$O$_4$ ($x$ = 0.20, 0.30, 0.50, 0.75) solid solutions were analysed using the reverse Monte Carlo method with the evolutionary algorithm (RMC/EA) approach, implemented in the EvAX code \cite{Timoshenko2012rmc,Timoshenko2014,evax2017}. 
Note that the use of the evolutionary algorithm as an optimization method significantly reduces the simulation time \cite{Timoshenko2014}. A detailed description of the RMC/EA method and its application to pure CuMoO$_4$ and CuMo$_{0.90}$W$_{0.10}$O$_4$ solid solution can be found in \cite{Jonane2019b,Jonane2020a}.  

To perform RMC/EA simulations, an initial model of the material structure should be constructed first. Here, the diffraction data reported in  \cite{Wiesmann1997,Schwarz2007} were used to build the structural model of CuMo$_{0.25}$W$_{0.75}$O$_4$. For other solid solutions ($x$=0.20, 0.30, 0.50), the lattice parameters corresponding to the CuMoO$_4$-III phase \cite{Wiesmann1997} and Wyckoff positions of atoms taken from the CuWO$_4$ structure \cite{Forsyth1991} were combined to build an initial unit cell. 

Next, the unit cell was enlarged 3$\times$3$\times$3 times, and W atoms were randomly substituted  with Mo atoms while maintaining the stoichiometric ratio of elements. 
The enlarged cell was additionally increased 2$\times$2$\times$2 times to obtain a supercell, which was employed as the RMC simulation box containing 2592 atoms in total. Such a large box was used to collect reasonable statistics for atoms that are the minority in the solid solution. 
Note that the RMC/EA simulations for pure CuWO$_4$ were performed  based on a structural model from \cite{Forsyth1991} employing a 5$\times$5$\times$5 large supercell with 1500 atoms.
The periodic boundary conditions (PBC) were imposed to avoid surface-related effects. 

Several RMC/EA simulations with different random substitutions and pseudo-random number sequences were performed to obtain better statistics. 
During the RMC/EA simulation, all atoms in the supercell were randomly displaced at each iteration with the maximum allowed displacement of 0.4~\AA\ with the goal to minimize the difference between the Morlet wavelet transforms (WTs) of the experimental and configuration-averaged (CA) EXAFS spectra \cite{Timoshenko2009wavelet}.  Such optimization criterion guarantees the agreement between experiment and theory simultaneously in $k$ and $R$ spaces. The number of atomic configurations (supercells) simultaneously considered in the EA algorithm was 32 \cite{Timoshenko2014}. In all simulations, 6000 iterations were made to ensure convergence.   

Theoretical CA-EXAFS spectra were calculated at each RMC/EA iteration for the given structure model using the ab-initio real-space multiple-scattering FEFF8.5L code  employing the complex exchange-correlation Hedin-Lundqvist potential \cite{Ankudinov1998,Rehr2000}, and the EXAFS amplitude reduction
factor $S^2_0$ was set to 1.0. Both single-scattering (SS) and multiple-scattering (MS) contributions were taken into account. 
MS contributions included all scatterings up to the 5th order, as higher-order scatterings contribute at large distances outside of the analysis range. A large number of scattering paths was reduced by grouping similar ones and including  into the analysis only important paths with the relative amplitude of corresponding partial contribution to the total EXAFS greater than 0.1--1\% \cite{Timoshenko2014}.

\begin{figure*}[t]
	\centering
	\includegraphics[width=\linewidth]{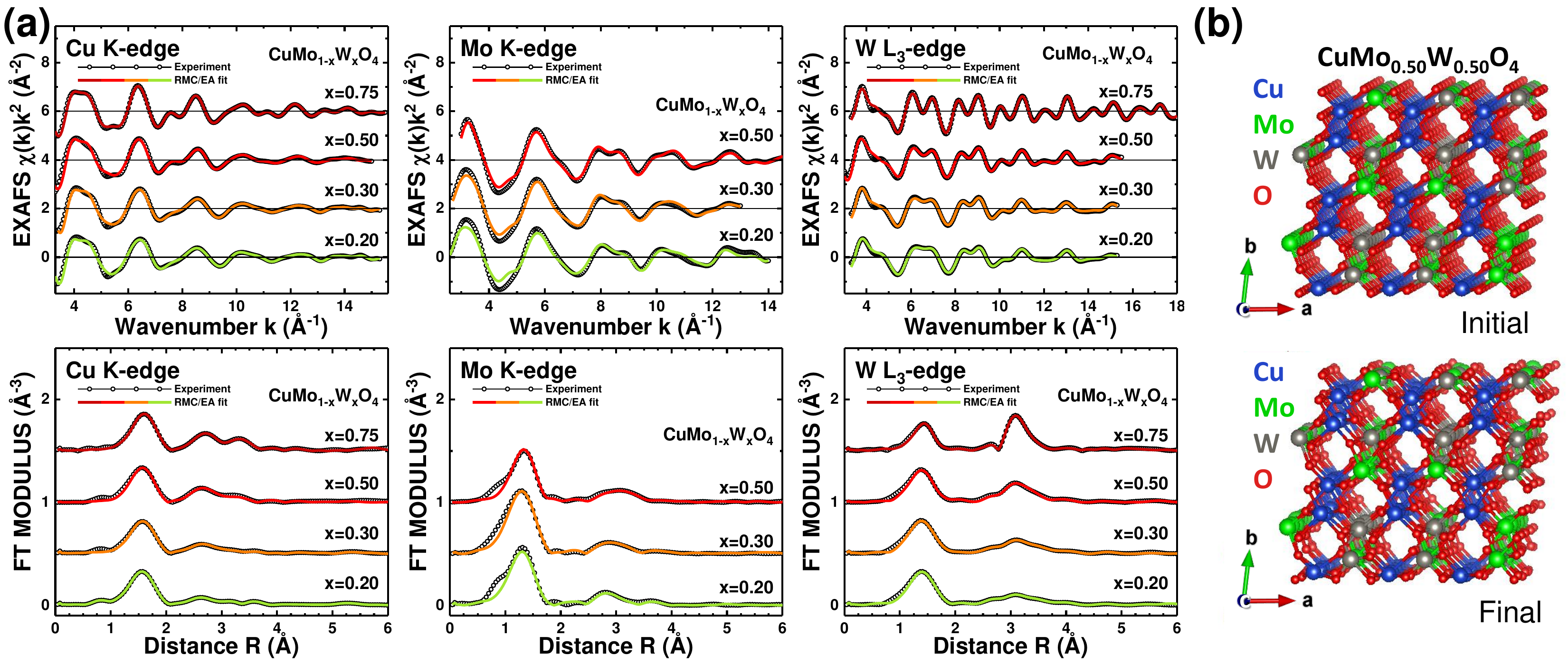}
	\caption{Results of RMC/EA calculations for the Cu K-edge, Mo K-edge and W L$_3$-edge in CuMo$_{1-x}$W$_x$O$_4$ (x=0.20, 0.30, 0.50 and 0.75) at room temperature. 
		(a) The experimental and calculated EXAFS spectra $\chi(k)k^2$ and their Fourier transforms (only modulus is shown). The FTs were calculated in the $k$-space range used in the fits and shown in the upper row. (b) Fragments of initial and final structural models for CuMo$_{0.50}$W$_{0.50}$O$_4$. }
	\label{fig3}
\end{figure*}

Examples of the obtained RMC/EA fits for all samples at three metal absorption edges (only two edges were used for CuMo$_{0.25}$W$_{0.75}$O$_4$) are shown in Fig.\ \ref{fig3}. As one can see, the calculated Cu and Mo K-edge and W L$_3$-edge EXAFS spectra, obtained from the multi-edge RMC/EA simulation, agree reasonably well with the corresponding experimental spectra. 

\begin{figure*}[t]
	\centering
	\includegraphics[width=\linewidth]{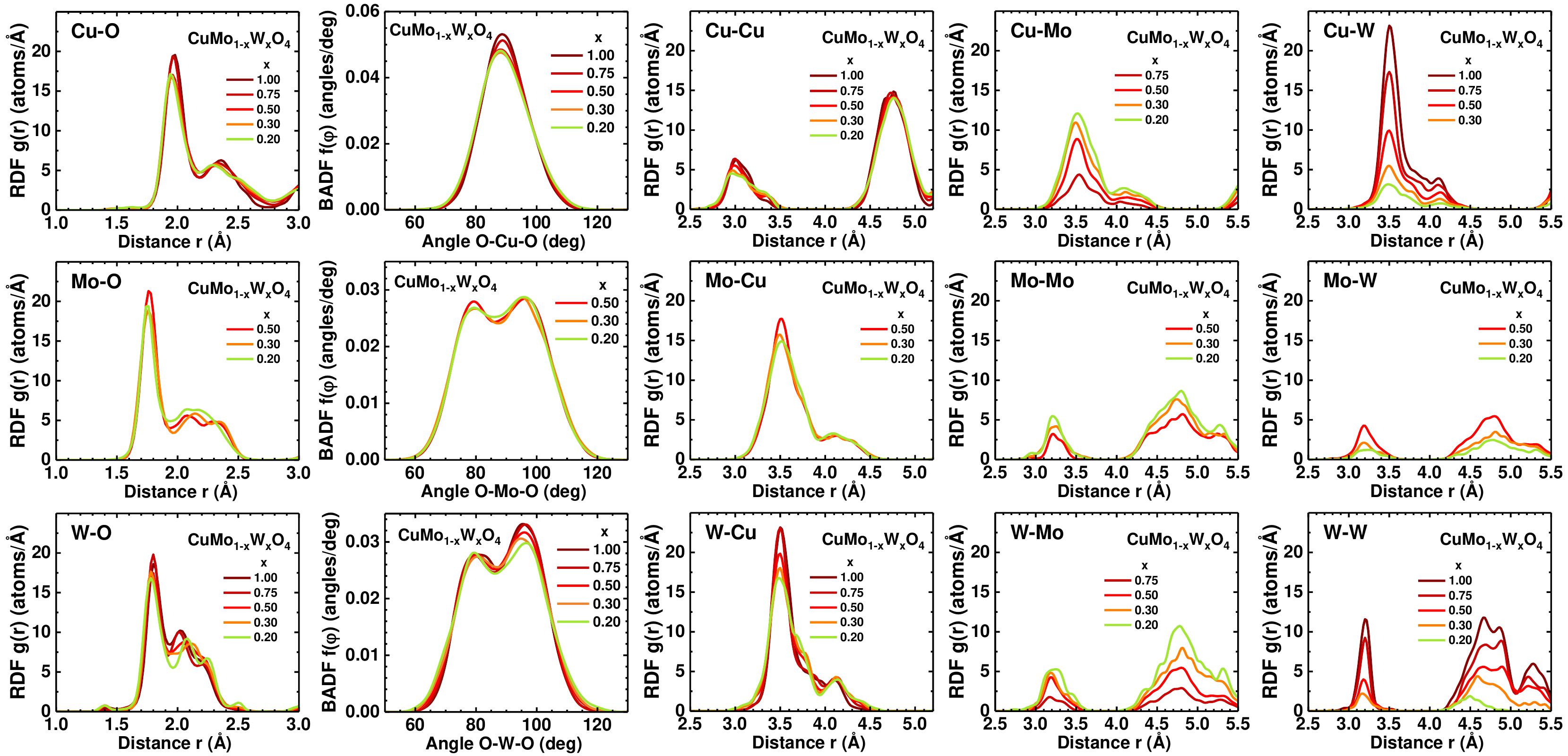}
	\caption{Partial radial distribution functions (RDFs) $g(r)$ and bond angle distribution functions (BADFs) $f(\varphi)$ in  CuMo$_{1-x}$W$_x$O$_4$ ($x$ = 0.20, 0.30, 0.50, 0.75, 1.00) obtained using RMC/EA calculations. }
	\label{fig4}
\end{figure*}

The final coordinates of atoms in the RMC simulation box were used to calculate partial radial distribution functions (RDFs) and bond angle distribution functions (BADFs) around absorbing atoms (Fig.\ \ref{fig4}).

\section{Results and discussion}

First, we describe the compositional dependencies of the experimental XANES and EXAFS spectra, as well as their FTs, while the structural changes obtained by the RMC/EA simulations will be discussed next.

The Cu K-edge XANES spectra of CuMo$_{1-x}$W$_{x}$O$_4$ solid solutions demonstrate relatively weak sensitivity to tungsten content (Fig.\ \ref{fig2}(a)). 
At the same time, the Cu K-edge EXAFS spectra (Fig.\ \ref{fig2}(d)) are more informative and reflect the gradual change in the copper local environment. In particular, there are pronounced changes in $R$-space above $\sim$2.1~\AA\ (Fig.\ \ref{fig2}(g)).
The difference in the shape of FTs in the range of 2.1--3.6~\AA\ in Fig.\ \ref{fig2}(g) reflects the dependence of the EXAFS signal on the composition of solid solutions, i.e., the substitution of Mo with W atoms. These two atoms are well separated in the Periodic Table and so have different scattering amplitudes that significantly affect their contribution to FTs even if they are located at close crystallographic distances. 

Molybdenum ions in $\alpha$-CuMoO$_4$ have tetrahedral coordination \cite{Wiesmann1997} which is responsible for the high intensity of the pre-edge peak at the Mo K-edge originating from the 1s(Mo)$\rightarrow$4d(Mo)+2p(O) transition (Fig.\ \ref{fig2}(b))  \cite{Jonane2018b,Jonane2019b,Pudza2021b}. Upon increasing tungsten content, the pre-peak amplitude decreases indicating a change in the local environment of molybdenum from tetrahedral to a more distorted one, which affects also the Mo K-edge EXAFS spectra (Fig.\ \ref{fig2}(e,h)). Indeed, the analysis of experimental Mo K-edge EXAFS spectra using the RMC/EA method (see below) suggests a change from the dominant tetrahedral to octahedral coordination above $x=0.10$.

The W L$_3$-edge XANES spectra are dominated by a strong white line peak at 10207~eV due to the dipole allowed transition 2p$_{3/2}$(W)$\rightarrow$5d(W)+2p(O). The shape of the white line contains information on the crystal field splitting of the W 5d-states \cite{Yamazoe2008,Jayarathne2014}, which is masked by a large value of the natural width ($\sim$4.5~eV \cite{KESKIRAHKONEN1974}) of the 2p$_{3/2}$(W) core level. As a result, the experimental XANES spectra do not show significant changes upon tungsten content increase due to the limited resolution. However, one can overcome this problem by performing resonant X-ray emission spectroscopy (RXES) experiments and obtaining high-energy resolution fluorescence detected (HERFD) XANES as reported in \cite{Pudza2021a}. Note that the analysis of the W L$_3$-edge HERFD-XANES allows one to identify the type of tungsten coordination distinguishing between tetrahedral and octahedral environments that can be used to determine the hysteretic behaviour of the temperature-induced structural phase transition between the $\alpha$ and $\gamma$ phases in CuMo$_{1-x}$W$_x$O$_4$  solid solutions with $x < 0.20$ \cite{Pudza2021a}. 

In our recent study \cite{Jonane2020a}, we found that an addition of 10~mol\% of tungsten to CuMoO$_4$ induces detectable local distortions and, while molybdenum ions maintain tetrahedral coordination as in pure $\alpha$-CuMoO$_4$, tungsten atoms tend to have a more distorted environment and promote an increase of the $\alpha$-to-$\gamma$ phase transition temperature. Indeed, the W L$_3$-edge EXAFS spectrum for $x=0.10$ differs from all other EXAFS
spectra (Fig.\ \ref{fig2}(f)). The difference is well visible in the $k$-space at $\sim$6.5~\AA$^{-1}$\  and in $R$-space at $\sim$3~\AA.
The spectra of brownish samples with $0.20 \leq x \leq 0.50$ show gradual change towards the spectra characteristic of the CuWO$_4$ phase where tungsten ions are octahedrally coordinated.  At an even larger tungsten content  $x$ = 0.75, the W L$_3$-edge EXAFS spectrum becomes close to that of pure CuWO$_4$. The transition to the CuWO$_4$ phase is accompanied by a significant increase of the second coordination shell peak at about 3~\AA\ in the FT (Fig.\ \ref{fig2}(i)): its amplitude continuously increases with an exchange of molybdenum atoms with tungsten ones. 

The detailed information on the structure of CuMo$_{1-x}$W$_x$O$_4$ solid solutions was obtained from the results of RMC/EA calculations (Fig.\ \ref{fig3}(a)). Using the RMC/EA method makes it possible to refine the structural model within the multiple scattering approach simultaneously for a set of EXAFS spectra measured at three (for $x$ = 0.20, 0.30, 0.50) or two (for $x$ = 0.75 and 1.0) absorption edges. Simultaneous analysis of absorption edges for all metals constituting our solid solutions is crucial for obtaining a reliable structural model as has been demonstrated previously for CuWO$_4$ in \cite{Timoshenko2014cuwo4}.
For example, we show in Fig.\ \ref{fig3}(b) fragments of the initial and final structural models for CuMo$_{0.50}$W$_{0.50}$O$_4$. A displacement of atoms from equilibrium positions due to an increase in the disorder after the RMC/EA simulation is well evident. The averaged partial RDFs and BADFs, calculated from the atomic coordinates of the several RMC/EA models, are shown in Fig.\ \ref{fig4} for the solid solutions and CuWO$_4$. 

In the CuMoO$_4$-III phase, which was used as a starting model to simulate the structure of the solid solutions ($x$=0.20, 0.30, 0.50), all metal ions are octahedrally coordinated by oxygen ones \cite{Wiesmann1997}. 
Due to the Jahn-Teller distortion, the CuO$_6$ octahedra are axially distorted with two groups of Cu--O bonds in an equilibrium structure: 4$\times$1.97~\AA, and 2$\times$(2.38--2.42~\AA) \cite{Wiesmann1997,Schwarz2007,Forsyth1991}. 
Molybdenum and tungsten ions are present in the 6+ oxidation state so that their coordination octahedra are subject to the second-order Jahn-Teller distortion with the off-center position of the metal ions \cite{Kunz1995}. 
It is visible that the MoO$_6$ octahedra are slightly stronger distorted than WO$_6$.
For both Mo and W ions, two main groups of metal-oxygen distances can be distinguished in the RDFs $g_{\rm Mo-O}(r)$ and $g_{\rm W-O}(r)$. The nearest group of three oxygen atoms is responsible for the first sharp peak in the RDFs and is located at about 1.8~\AA\ for both W and Mo. The second group of three oxygen atoms is quite broad and can be roughly  divided into two subgroups with 2 and 1 oxygen atoms. In the case of tungsten, two subgroups of oxygen atoms are located at the mean distances of about 2.0--2.1~\AA\ and 2.15--2.25~\AA, while in the case of molybdenum, both distances are slightly longer -- 2.05--2.15~\AA\ and 2.20--2.40~\AA.

Various types of distortions of metal-oxygen octahedra  are well reflected by the BADFs $f(\varphi)$ (Fig. \ \ref{fig4}). 
Axial distortion of CuO$_6$ octahedra results in a monomodal BADF O--Cu--O with an average angle value of $\varphi \approx $88$^\circ$. At the same time, off-center displacement of Mo or W atoms from the center of the octahedra leads to bimodal BADFs O--Mo--O and O--W--O with maxima at $\varphi \approx $80$^\circ$ and 96$^\circ$.         

Proper account for the multiple-scattering and disorder effects in the calculated EXAFS spectra  allows a reliable analysis of the contributions from the outer coordination shells. 
In Fig.\ \ref{fig4}, the partial RDFs for metal--metal atom pairs are shown up to 5.0--5.5~\AA. 
In the structure of CuMo$_{1-x}$W$_x$O$_4$ solid solutions with $x \geq 0.20$ (Fig.\ \ref{fig3}(b)), the 
metal--oxygen distorted octahedra of one type share edges and form zigzag chains along the $c$-axis \cite{Schwarz2007}. The zigzag chains are arranged in alternating layers that are perpendicular to the $a$-axis direction \cite{Schwarz2007}.

The RDFs $g_{\rm Cu-Cu}(r)$ include contributions from two nearest copper atoms located 
within the same chain of CuO$_6$ octahedra (the peak at $\sim$3.0~\AA) and six more distant copper atoms (four in the neighbouring chains and two in the same chain), giving the origin of the peak at $\sim$4.7~\AA.

Eight copper atoms located in the four neighbouring chains of CuO$_6$ octahedra contribute to  the RDFs $g_{\rm Mo/W-Cu}(r)$. The change in the composition of solid solutions leads to some small modifications in the structure resulting in the broadening 
of the main peak at $\sim$3.5~\AA\ at low tungsten content ($x \leq 0.50$).

The RDFs $g_{\rm Cu/Mo/W-Mo/W}(r)$ show a clear dependence on the number of Mo/W atoms varying with the composition. Upon increasing tungsten content, the amplitude of the peaks becomes  larger in the RDFs $g_{\rm Cu/Mo/W-W}(r)$ but decreases in the RDFs $g_{\rm Cu/Mo/W-Mo}(r)$. The ability of molybdenum atoms to be in a locally distorted environment extends to the outer shells so that the RDFs $g_{\rm Mo-Cu/Mo}(r)$ are stronger broadened than the RDFs $g_{\rm W-Cu/W}(r)$.  

Thus, our results show that both Mo and W ions are octahedrally coordinated by oxygen ions in solid solutions with large tungsten content ($x \gtrsim  0.2$). 
However, previous studies of CuMo$_{1-x}$W$_x$O$_4$  solid solutions revealed that at room temperature for $x \leq 0.15$, tungsten ions co-exist in the octahedral and tetrahedral environment \cite{Pudza2021a}, while molybdenum ions  have dominant tetrahedral coordination \cite{Gaudon2007a,Gaudon2008,Jonane2020a}. This means that tungsten atoms have a decisive influence on the structure of CuMo$_{1-x}$W$_x$O$_4$ solid solutions.  

\begin{figure*}[t]
	\centering
	\includegraphics[width=\linewidth]{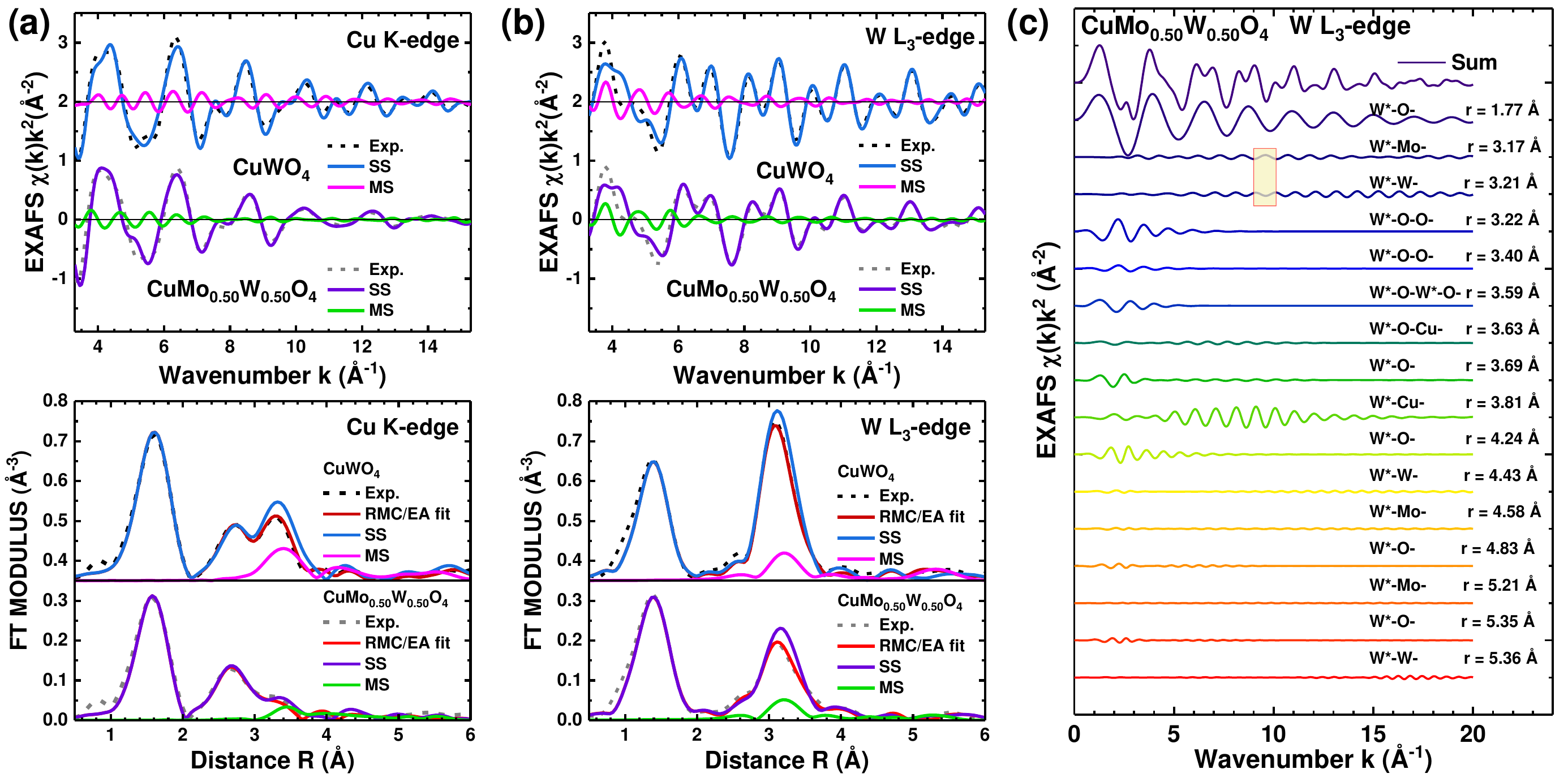}
	\caption{Contributions of different scattering paths into the Cu K-edge and W L$_3$-edge EXAFS spectra of CuWO$_4$ and  CuMo$_{0.50}$W$_{0.50}$O$_4$: Cu K-edge (a) and W L$_3$-edge (b) EXAFS spectra and corresponding FTs calculated for the $k$-space range 3.3--15.3~\AA$^{-1}$. SS and MS contributions are calculated for a structure configuration obtained in RMC/EA simulation. Contributions of selected scattering paths are shown at the W L$_3$-edge for  CuMo$_{0.50}$W$_{0.50}$O$_4$ structure (c). Mismatch of phases in W--Mo and W--W scattering paths is highlighted with a rectangle.   }
	\label{fig5}
\end{figure*}

Finally, we would like to draw attention to the origin of large variations of some peaks in FTs, well observed in the range of 3--4~\AA\  at the Cu K-edge and W L$_3$-edge in Fig.\ \ref{fig5}(a,b). Here the contributions of the SS and MS processes are shown for two samples -- CuWO$_4$ and CuMo$_{0.50}$W$_{0.50}$O$_4$.
Additionally, a set of the partial  W L$_3$-edge EXAFS contributions for selected photoelectron scattering paths is presented for CuMo$_{0.50}$W$_{0.50}$O$_4$ in Fig.\ \ref{fig5}(c). 
The relatively small amplitude of the MS oscillations compared to the SS ones in the $k$-space  (Fig.\ \ref{fig5}(a,b)) looks deceiving because the SS signal from the first coordination shell dominates the total EXAFS spectrum. However, when compared in $R$-space, the MS contribution reaches from 20--30\% (at the W L$_3$-edge) up to 50--60\% (at the Cu K-edge) of the SS one in the range of 3--4~\AA\ and even more at larger distances. Therefore, the MS contribution should not be ignored in the RMC simulation to get a reliable structural model.

In CuMo$_{1-x}$W$_x$O$_4$ solid solutions having close crystallographic structures and, thus, close geometry of scattering paths, the changes in the EXAFS spectra are caused by the different contributions of molybdenum and tungsten atoms, due to the large difference in their atomic masses and electron configurations. The atomic mass of W (183.84~amu) is almost twice larger than that of Mo (95.95~amu) which should lead to the difference in the amplitude of their thermal vibrations and, thus, in the damping of EXAFS oscillations due to the difference in the mean-square relative displacement (MSRD) factors. Also, the molybdenum ($Z$(Mo)=42) and tungsten ($Z$(W)=74) atoms are well separated in the Periodic Table, so they will make different contributions to the EXAFS spectra due to the difference in their scattering amplitudes. Note that the latter effect will influence only FTs but not RDFs.
Both effects are responsible for a strong reduction of the peak at $\sim$3.1--3.2~\AA\ 
in the Cu K-edge and W L$_3$-edge FTs when 50\% of tungsten atoms are substituted with molybdenum ones (Fig.\ \ref{fig5}(a,b)). 

It is also interesting that the W L$_3$-edge EXAFS contributions from W--Mo and W--W scattering paths with $r \sim 3.2$~\AA\ have nearly opposite phases (Fig.\ \ref{fig5}(c)) leading to destructive interference. In addition, damping of the EXAFS oscillations is larger for the W--Mo scattering path due to a wider W--Mo distribution (MSRD $\sigma^2_{\rm W-Mo} \approx 0.0163$~\AA$^2$) than for W--W (MSRD $\sigma^2_{\rm W-W} \approx 0.0049$~\AA$^2$), which is associated with larger static distortions of the molybdenum environment (Fig.\ \ref{fig4}).

\section{Conclusion}

Multi-edge X-ray absorption spectroscopy was used to study the local atomic structure in CuMo$_{1-x}$W$_x$O$_4$ ($x$ = 0.20, 0.30, 0.50, 0.75) solid solutions at room temperature. The structural models for each composition  were obtained from the simultaneous analysis of EXAFS spectra measured at several (Cu K-edge, Mo K-edge and W L$_3$-edge) absorption edges by the reverse Monte Carlo method. By minimizing the difference between the Morlet wavelet transforms of the experimental and theoretical EXAFS spectra, good agreement was achieved in both $k$ and $R$ spaces, which indicates the reliability of the obtained structural models.    

It was shown that multiple-scattering effects play an important role in the analysis of CuMo$_{1-x}$W$_x$O$_4$  solid solutions and should be taken into account to describe accurately the contributions from the outer coordination shells. 

A detailed analysis of the partial radial distribution functions (RDFs) $g(r)$ around absorbing metal atoms and bond angle distribution functions (BADFs) $f(\varphi)$ gave  information on the degree of distortion of the coordination shells and its dependence on the composition. 

Molybdenum and tungsten atoms are octahedrally coordinated by oxygen atoms in all studied
solid solutions; however, MoO$_6$ octahedra are somewhat more distorted than WO$_6$.  
For both metals, the distorted octahedra consist of three short and three long metal--oxygen bonds, and the group of the nearest three oxygen atoms has narrow distribution. 

The ability of molybdenum atoms to adopt a locally distorted environment allows them to adjust to the solid solution structure determined by tungsten-related sublattice. Only at low tungsten content of about $x$ = 0.15 \cite{Pudza2021a},  a transition  from CuWO$_4$-type structure with octahedral coordination of tungsten to a mixture of $\alpha$ and $\gamma$-CuMoO$_4$-type structures occurs for  $0.075 < x <0.15$, followed by a transition  to $\alpha$-CuMoO$_4$-type structure  with tetrahedral coordination of molybdenum for $0 < x < 0.075$ \cite{Gaudon2007a,Gaudon2008}.

\section*{Data availability}
Data associated with this article are available upon reasonable request to the authors.

\section*{Declaration of Competing Interest}
The authors declare that they have no known competing financial interests or personal relationships that could have appeared to influence the work reported in this paper.

\section*{Acknowledgements}

I.P. and A.K. would like to thank the support of the Latvian Council of Science project No. lzp-2019/1-0071. I.P. acknowledges the L'OR{\'E}AL Baltic ``For Women In Science Program'' with the support of the Latvian National Commission for UNESCO and the Latvian Academy of Sciences.	
The experiment at the Elettra synchrotron was performed	within project No. 20150303.
Institute of Solid State Physics, University of Latvia as the Center of Excellence has received funding from the European Union's Horizon 2020 Framework Programme H2020-WIDESPREAD-01-2016-2017-TeamingPhase2 under grant agreement No. 739508, project CAMART2.`

\newpage


\end{document}